\documentclass[%
reprint,
superscriptaddress,
showpacs,
 amsmath,amssymb,
 aps,
 prl,
]{revtex4-1}

\usepackage{graphicx}
\usepackage{dcolumn}
\usepackage{bm}
\usepackage{amsmath}
\usepackage{autobreak}
\usepackage{hyperref}
\usepackage[mathlines]{lineno}
\usepackage{natbib}
\usepackage{multirow}
\usepackage{xspace}
\usepackage{float}
\usepackage{diagbox}
\usepackage{subfigure}
\usepackage{booktabs}
\usepackage{array}
\usepackage{upgreek}

\begin{document}

\preprint{APS/123-QED}

\title{Novel Search for absorption of ultra-light fermionic dark matter \\on electron with cosmic ray boost}

\author{X.~P.~Geng}
\email{Corresponding author: gxp18@tsinghua.org.cn}
\affiliation{Institute of Applied Physics and Computational Mathematics, Beijing 100094, China}

\author{B.~Zhong}
\affiliation{Institute of Applied Physics and Computational Mathematics, Beijing 100094, China}

\author{Z.~H.~Zhang}
\affiliation{Beijing Normal University at Zhuhai, Zhuhai 519087, China}

\date{\today}

\begin{abstract}

The search for the absorption of fermionic dark matter (DM) on electron is hindered by a partial or complete loss of sensitivity for the mass of DM ($m_\chi$) below $\mathcal{O} (10)$ keV. We introduce a novel search using a small but high-energy portion of DM boosted by cosmic rays, and the loss of sensitivity is circumvented because the subsequent absorption within the detector imparts to the target electron not only $m_\chi$ but also substantial $T_\chi$. The one-sided 90\% confidence level constraints on ($\sigma_e v_\chi$) are derived with the XENONnT public dataset for $m_\chi$ $\in$ (1$\times$10$^{-5}$--1) keV which has not been explored yet. Our results have placed constraints on DM at the lowest $m_\chi$ currently achievable among the DMDD experiments utilizing ionizing signals from DM interactions within the detectors.

\end{abstract}

\maketitle

\emph{Introduction.}—
A multitude of astronomical and cosmological observations strongly imply the presence of dark matter (DM, denoted as $\chi$), whose enigmatic nature has persisted as one of the most elusive and intriguing scientific challenges for decades~\cite{DM1,DM2,DM-RESEARCH1,DM-RESEARCH2,DM-RESEARCH3}. Numerous prominent DM candidates, including weakly interacting massive particles~\cite{XENON-DM,SuperCDMS-DM,LUX-ZEPLIN-DM,PandaX-DM,DarkSide-DM,SENSEI-DM,CDEX-DM}, dark photon DM~\cite{XENONnT-ALP-DP,XENON1T_ER2020,XENON1T-ALP-DP,RFC-DP,CDEX-10-DP,CDEX-10-DP2,DARMIC-DP,SuperCDMS-DP,XMASS-ALP-DP,TEXONO-DP}, and axion-like particle DM~\cite{XENONnT-ALP-DP,XENON1T-ALP-DP,CAST-ALP,CAST-ALP2,Amplifier-ALP,ferromagnets-ALP,LUX-ALP,EDELWEISSIII-ALP,CDEX-1B-ALP,Majorana-ALP,PandaX-II-ALP,SuperCDMS-ALP,XMASS-ALP-DP}, have been the focus of vigorous experimental exploration, yet the absence of definitive results has prompted the development of alternative DM models~\cite{brem1,brem2,DM-absorption1,migdal1}.
\par
Recently, a newly presented interaction scenario involving the absorption of fermionic dark matter by an electron target ($\chi+e^- \rightarrow{} \nu+e^-$) has garnered attention and been subjected to investigation within the research community~\cite{DM-absorption-electron-PandaX-4T,absorption,revisiting,CDEX-absorption}. Within this specific scenario, the absorption of dark matter is accompanied by the emission of an outgoing neutrino, and the target electron can acquire kinetic energy from both the mass of DM ($m_\chi$) and the kinetic energy of DM ($T_\chi$). The DM detectors can register the corresponding energy deposited, providing an opportunity to search for this scenario in the dark matter direct detection (DMDD). 
\par
In previous studies of the scenario, the Standard Halo Model (SHM)~\cite{SHM1,SHM2} is adopted, wherein DM possessing a typical velocity $v_\chi\sim$ 10$^{-3} c$ results in a negligible $T_\chi$ compared to $m_\chi$~\cite{DM-absorption-electron-PandaX-4T, absorption, revisiting,CDEX-absorption}. The latest results from PandaX-4T~\cite{DM-absorption-electron-PandaX-4T} have reached a detection lower limit of $m_\chi = $ 10 keV and its ``optimal'' $m_\chi$ region lies on $m_\chi$ $\in$ (40--100) keV with corresponding ($\sigma_e v_\chi$) $<$ 1$\times$10$^{-49}$cm$^2$ for axial-vector interaction. While below the ``optimal'' $m_\chi$ region, the direct detection sensitivity to DM diminishes swiftly due to the reduction in electronic recoil energy which cannot exceed $m_\chi$. Furthermore, if $m_\chi$ falls below the detection threshold, the detector will completely lose its capability to detect this scenario.

\par
Fortunately, it has come to light that numerous processes can accelerate a subdominant component of DM to a much higher $T_\chi$. Primary potential avenues include: (1) DM particles collide with high-energy Galactic cosmic rays (CR)~\cite{novel,CRDM_PandaX,CRDM_CDEX-10,CRDM_ICECUBE}. (2) DM particles engage in scattering interactions with rapidly moving nuclei or electrons inside the Sun (solar reflection)~\cite{solar_reflection,solar_reflection2,solar_reflection_CDEX}. (3) DM particles from the evaporation of Primordial Black Holes can possess higher energy~\cite{PBH,PBH_CDEX}. 
\par
We consider a secondary yet inevitably present fraction of the DM flux where DM receives an impetus from the aforementioned CR and subsequently attains a significantly elevated $T_\chi$, this specific part of DM is referred to as CRDM. The CRDM will go through the aforementioned absorption process with both $m_\chi$ and notable $T_\chi$, leading to a heightened deposited energy registered in the detector. As a consequence, the detectable range of the $m_\chi$ can be extended into the ultra-light region, enabling the exploration of previously inaccessible areas of DM parameter space (theoretically extending to $m_\chi \rightarrow{} 0$). 
\par
In this letter, leveraging the kinetic enhancement provided by CR, we employ the electronic recoil data from XENONnT~\cite{XENONnT_EPJC2024} which is one of the most state-of-the-art DMDD experiments, to perform a search for the absorption of ultra-light fermionic DM on electron, and the constraints of this scenario are derived with $m_\chi$ $\in$ (1$\times$10$^{-5}$--1) keV. The boost of DM from CR is presumed to occur via spin-independent (SI) scattering in this work, this assumption is made as the paper does not intend to delve into the mechanisms underlying the production of CRDM.

\emph{CRDM flux.}—
CRs in the vicinity of Earth, which are assumed to originate from supernova remnants, are high-energy particles consisting of protons, electrons, and an assortment of heavier nuclei such as $^4$He~\cite{CR1,CR2}. 
Electrically charged CRs interact with Galactic magnetic fields during their propagation. These interactions cause the CR trajectories to undergo numerous scatterings due to the irregular and turbulent nature of the magnetic fields. This process effectively randomizes their original directions, leading to the isotropic distribution observed on Earth~\cite{CR_mag}. The Local Interstellar Spectra (LIS) of CR proton and $^{4}$He, the two primary components, can be evaluated by Eq.~\ref{eq_CR_I}.
\begin{equation}
\label{eq_CR_I}
\frac{dI}{dR} \times R^{2.7} = 
\begin{cases} 
\sum_{i=0}^{5} a_i R^i, & R \leq 1 \, \text{GV} \\ 
b + \frac{c}{R} + \frac{d_1}{d_2 + R} + \frac{e_1}{e_2 + R}\\ + \frac{f_1}{f_2 + R} + gR, & R > 1 \, \text{GV}
\end{cases}
\end{equation}
where $R$ is the magnetic rigidity of the specific particle, the rest parameters are listed in Table.~\ref{tab_CR_parameters}.

\begin{table}[htbp]

\centering
\caption{The efficiencies in Eq.~\ref{eq_CR_I} describing the differential intensities of CR proton and $^4${He}~\cite{CR_parameters1,CR_parameters2}.}
\label{tab_CR_parameters}
\begin{tabular}{lcccccccc}\hline
	&$a_0$ &$a_1$ &$a_2$ &$a_3$ &$a_4$ &$a_5$ &$b$ &c \\\hline
	p &94.1 &-831 &0 &16700 &-10200 &0 &10800 &8500\\\hline
	$^4$He &1.14 &0 &-118 &578 &0 &-87 &3120 &-5530\\\hline
	&$d_1$ &$d_2$ &$e_1$ &$e_2$ &$f_1$ &$f_2$ &$g$ &\\\hline
	p &-4230000 &3190 &274000 &17.4 &-39400 &0.464 &0&\\\hline
	$^4$He &3370 &1.29 &134000 &88.5 &-1170000 &861 &0.03&\\\hline
    
\end{tabular}
\end{table}

The LIS of CR flux as a function of kinetic energy $T$ is expressed by $\frac{d\Phi^{\rm LIS}}{dT} = 4\uppi\frac{dI}{dR}\frac{dR}{dT}$, where $\frac{dR}{dT}$ should be calculated with account for relativity shown in Eq.~\ref{eq_DRDT}.
\begin{align}
\label{eq_DRDT}
\frac{dR}{dT} = \frac{1}{Q}\frac{T+m_i}{T^2 + 2Tm_i},
\end{align}
where $Q$ is the electric charge of CR particle $i$, $m_i$ is the mass of CR particle $i$.
\par
DM particles can be considered at rest compared to CR particles traveling at relativistic velocities. The kinetic energy transferred to a DM particle $T_\chi$ from the CR particle in a single collision is described in Eq.~\ref{eq_transfer}.
\begin{align}
\label{eq_transfer}
T_\chi &= T_\chi^{\text{max}} \frac{1 - \cos \theta_s}{2},\nonumber\\
T_\chi^{\text{max}} &= \frac{T_i^2 + 2 m_i T_i}{T_i + (m_i + m_\chi)^2/(2 m_\chi)},
\end{align}
where $T_i$ is the kinetic energy of the incident CR particle $i$, $m_\chi$ is the mass of DM, $\theta_s$ is the  scattering angle between DM and CR particle $i$ in the center-of-mass reference frame. Conversely, the minimum kinetic energy of the CR particle required to impart a DM kinetic energy of 
$T_\chi$ is given by Eq.~\ref{eq_T_min}. 
\begin{equation}
  T_i^{\text{min}} = \left( \frac{T_{\chi}}{2} - m_i \right) \left( 1 \pm \sqrt{1 + \frac{2 T_{\chi} \left( m_i + m_{\chi} \right)^2}{m_{\chi} \left( 2 m_i - T_{\chi} \right)^2}} \right),
  \label{eq_T_min}
\end{equation}
where + (-) corresponds to $T_\chi > 2m_i$ ($T_\chi < 2m_i$).

\par
The differential CRDM flux is derived by Eq.~\ref{eq_CRDM1}, with integration over all the CR proton and $^{4}$He along the Line of Sight (LOS).
\begin{equation}
    \frac{d\Phi_\chi}{dT_\chi} = \sum_i \int \frac{d\Omega}{4\uppi} \int_{\text{LOS}} dl \int_{r_i^{\min}}^\infty \frac{\rho_\chi(r)}{m_\chi} \frac{d\sigma_{\chi i}}{dT_\chi} \frac{d\Phi_i(r)}{dT_i} dT_i,
    \label{eq_CRDM1}
\end{equation}
where $\Omega$ is the solid angle, $\rho_\chi(r)$ is the DM density distribution which the Navarro-Frenk-White (NFW) profile~\cite{DM_distribution1,DM_distribution2} is adopted, $\frac{d\Phi_i(r)}{dT_i}$ is the CR flux spatial distribution of the CR particle $i$, treated as homogeneous in this work~\cite{novel}. The differential cross-section of DM and CR particle $\frac{d\sigma_{\chi i}}{dT_\chi}$ is shown in Eq.~\ref{eq_cross_section}
\begin{equation}
\frac{d\sigma_{\chi i}}{dT_\chi} = \frac{\sigma^{\rm SI}_{\rm \chi N}}{T^{\text{max}}_\chi} A_i^2 \left( \frac{\mu_{\chi i}}{\rm \mu_{\chi N}} \right)^2 G^2_i(Q^2),
\label{eq_cross_section}
\end{equation}
where $\sigma^{\rm SI}_{\rm \chi N}$ is the zero-momentum transferred DM-nucleon cross section for SI scattering, $A_i$ is the mass number of CR particle $i$, $\mu_{\chi i}$ and $\rm \mu_{\chi N}$ are the reduced masses of DM-$i$ and DM-nucleon, respectively. $G(Q^2)$ is the form factor with $Q^2 = 2m_\chi T_\chi$, for proton and $^4$He, the dipole form factor $G_i(Q^2) = 1 / (1 + Q^2 / \Lambda_i^2)^2$ is adopted, where $\Lambda_p \simeq 770$ MeV and $\Lambda_{\rm {^4}He} \simeq 410$ MeV~\cite{formfactor}. 
\par
\begin{figure} [htb]
	\centering
	\includegraphics[width=\linewidth]{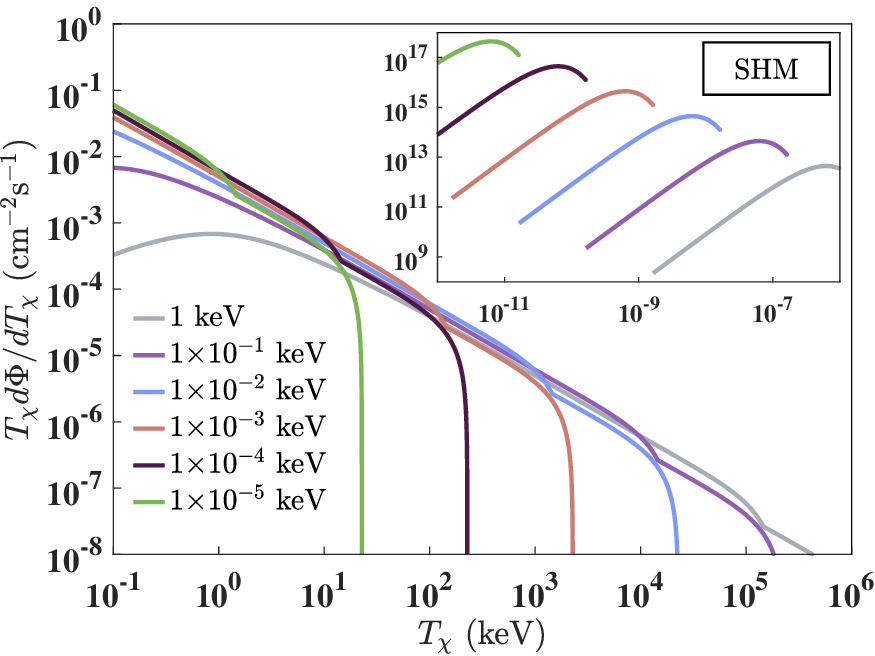} 
	\caption{CRDM fluxes for various $m_\chi$ with $\sigma_{\rm \chi N}^{\rm SI} = 10^{-31}$ cm$^2$ and $D_{\text{eff}} = 0.997$ kpc. Inset: The DM fluxes under the SHM for the same $m_\chi$, the axes of the inset are identical to those of the main figure.}
	\label{fig1_CRDM} 
\end{figure}

Combining the related items above, we can obtain $\frac{d \Phi_{\chi}}{d T_{\chi}}$ shown in Eq.~\ref{eq_CRDM2}.
\begin{align}
   \frac{d \Phi_{\chi}}{d T_{\chi}} &= D_{\text{eff}} \frac{\rho_{\chi}^{\text{local}}}{m_{\chi}} \nonumber\\
   &\times \sum_i \sigma_{\rm \chi N}^{\rm SI}A_i^2(\frac{\mu_{\chi i}}{\rm \mu_{\chi N}})^2G_i^{2}(2m_{\chi} T_{\chi}) \int_{T_i^{\min}}^{\infty} dT_i \, \frac{d \Phi_i^{\text{LIS}} / dT_i}{T_{\chi}^{\max}(T_i)},
   \label{eq_CRDM2}
\end{align}
where $\rho_\chi^{\text{local}}\sim0.3$ GeV/cm$^{3}$ is the local density of DM, $D_{\text{eff}}$ is the spatial extent effectively contributing to CR and DM interactions, dependent on their distributions in the Milky Way. With the NFW profile for the DM distribution and the homogeneous CR spatial distribution, $D_{\text{eff}} = \int\frac{d\Omega}{4\uppi} \int_{\rm LOS}\frac{\rho_\chi(r)}{\rho_\chi^{\text{local}}}dl$, the LOS is set to 1 kpc to obtain a conservative result, yielding $D_{\text{eff}}$ = 0.997 kpc in this work. The CRDM fluxes for various $m_\chi$ are shown in Fig.~\ref{fig1_CRDM}, with the DM fluxes under the SHM for comparison.

\emph{Signal model.}—
The fermionic CRDM-electron absorption can be induced by vector (V) and axial-vector (A) operators~\cite{absorption} shown in Eq.~\ref{eq_VA_operators}
\begin{align}
    \label{eq_VA_operators}
	\mathcal{O}_{e\nu\chi}^{V} &= \frac{1}{\Lambda^2}(\bar{e}\gamma_\mu e)(\bar{\nu}_L\gamma^\mu \chi_L),\nonumber\\
    \mathcal{O}_{e\nu\chi}^{A} &= \frac{1}{\Lambda^2}(\bar{e}\gamma_\mu \gamma_5 e)(\bar{\nu}_L\gamma^\mu \chi_L),
\end{align} 
where the Standard Model neutrino is taken, 1/$\Lambda^2$ is the Wilson coefficient with a dimension of (mass)$^{-2}$. 
\par
Momentum conservation gives Eq.~\ref{eq_momentum_conservation} describing the outgoing momentum of the neutrino.
\begin{align}
\label{eq_momentum_conservation}
p_\nu=|p_\chi-q|=\sqrt{p_\chi^2+q^2-2m_\chi v_\chi q\cos\theta_{p_\chi q}},
\end{align}
where $p_\chi$ is the momentum of the DM particle, $q$ is the momentum transferred to the target electron, $\theta_{p_\chi q}$ is the angle between $p_\chi$ and $q$ under the lab frame. In combination with the minuscule mass of the active neutrino~\cite{neutrino}, we can obtain Eq.~\ref{eq_energy_conservation} via energy conservation.
\begin{align}
    \label{eq_energy_conservation}
	m_\chi + T_\chi - \sqrt{p_\chi^2 +q^2-2p_\chi q\theta_{p_\chi q}}= |E^{nl}_B| + E_R,
\end{align}
where $E_R$ is the recoil energy, $|E^{nl}_B|$ is the binding energy of the atomic electron with quantum number ($n$,$l$)~\cite{binding_energy}. The deposited energy in the detector $E_{det}$ comprises both the binding energy and the recoil energy, $E_{det} = |E^{nl}_B| + E_R$. Correspondingly, $q$ is derived in Eq.~\ref{eq_q}.
\begin{align}
\label{eq_q}
q = p_\chi {\rm cos}\;\theta_{p_\chi q}\pm\sqrt{p_\chi^2({\rm cos^2}\;\theta_{p_\chi q} - 1) + (T_\chi + m_\chi - E_{det})^2},
\end{align}
where + (-) corresponds to ${\rm cos}\;\theta_{p_\chi q} \leq 0$ (${\rm cos}\;\theta_{p_\chi q} > 0$).
\par
The expected differential event rate of fermionic DM-electron absorption is shown in Eq.~\ref{eq_R}.
\begin{align}
\label{eq_R}
	\frac{dR
    }{dE{_{det}}} &= N_T\sum_{nl}\int v_\chi\frac{d\Phi_\chi}{dT_\chi} \frac{q}{64\uppi m_e^2m_\chi(E_{det} - |E_{B}^{nl}|)}\nonumber\\ 
    & \times |M(q)|^2|f_{ion}^{nl}(k^\prime,q)|^2dT_\chi,
\end{align}
where $N_T$ is the number of targets per unit mass with all ($n$,$l$) shell bound electrons included, $m_e$ is the electron mass, $|f_{ion}^{nl}(k^\prime,q)|^2$ with $k^\prime = \sqrt{(E_{det} - |E^{B}_{nl}| + m_e)^2 - m_e^2}$ is the ionization form factor calculated by $DarkART$~\cite{DarkART1,DarkART2}. $|M(q)|^2$ is the scattering matrix element shown in Eq.~\ref{eq_M} with different operators.
\begin{align}
\label{eq_M}
	|M^{(V,A)}(q)|^2 = (1,3)\times \frac{16\uppi m_e^2q}{m_\chi}\;(\sigma_e v_\chi),
\end{align}
where $\sigma_e$ is the cross section of free electron-DM scattering and can be formulated as $\sigma_e \equiv (m_\chi^2/4\uppi v_\chi \Lambda^4)$. 
\par
\begin{figure} [htb]
	\centering
	\includegraphics[width=\linewidth]{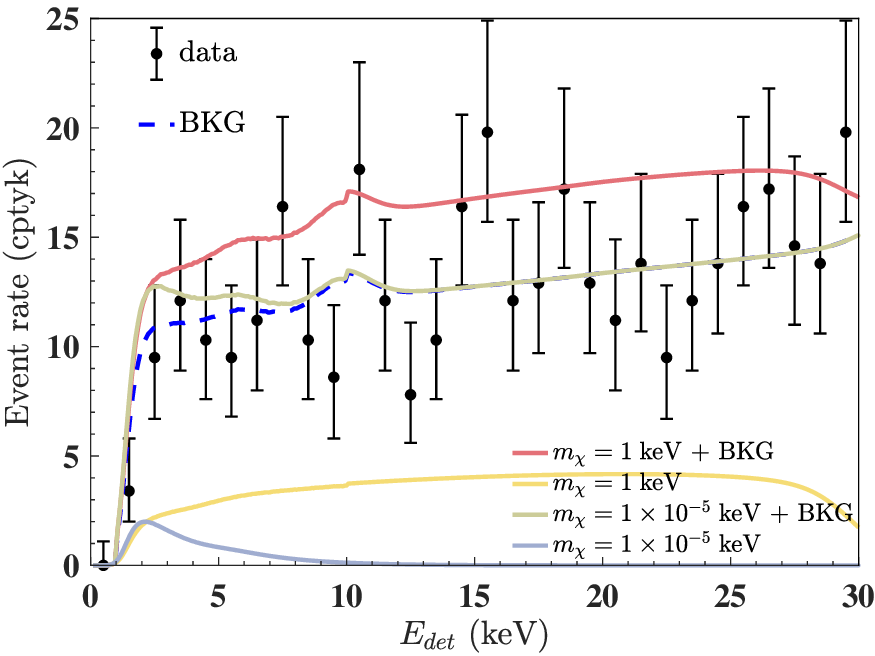} 
	\caption{Expected event rates of fermionic CRDM-electron absorption for ($m_\chi$, $\sigma_ev_\chi$) = (1 keV, 2$\times$10$^{-33}$ cm$^2$) and (1$\times$10$^{-5}$ keV, 2$\times$10$^{-41}$ cm$^2$), with $D_{\rm eff} = 0.997$ kpc and $\sigma_{\rm \chi N}^{\rm SI} = 10^{-31}$ cm$^{2}$. The electronic recoil data from XENONnT public dataset is plotted as points with error bars, the blue line represents XENONnT background model (BKG) extracted from Ref.~\cite{XENONnT_ER2022}.}
	\label{fig2_Signal+BKG}
\end{figure}

The expected event rates of fermionic CRDM-electron absorption induced by V operator in the XENONnT detector with $m_\chi$ = 1 / 1$\times$10$^{-5}$ keV are shown in Fig.~\ref{fig2_Signal+BKG} as an example, along with the XENONnT electronic recoil data. 
\par
\par
\emph{Exclusion result.}—
We apply the minimum-$\chi^{2}$ analysis in this work, where the $\chi^2$ is defined in Eq.~\ref{eq_chi2}.
\begin{equation}
\chi^2 = \sum_i \frac{\left[n_i-\left(S_i+B_i\right)\right]^2}{\Delta^2_i},
\label{eq_chi2}
\end{equation}
where $n_{i}$ is the measured count rate in the detector at the $i$-th energy bin, $\Delta_i$ is the uncertainty at the $i$-th energy bin with both the statistical and systematic components, $S_i$ is the expected count rate within the signal model at the $i$-th energy bin, and $B_i$ is the background model of XENONnT~\cite{XENONnT_ER2022} at the $i$-th energy bin.
\par
\begin{figure} [htb]
	\centering
	\includegraphics[width=\linewidth]{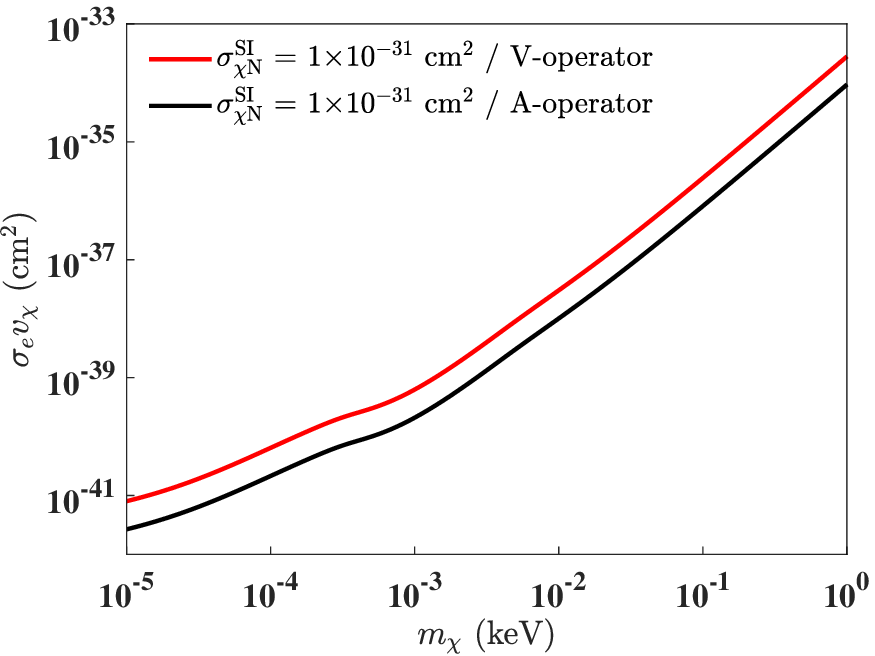} 
	\caption{One-sided 90\% CL constraint on ($\sigma_ev_\chi$) for the V operator and A operator imposed by the XENONnT experiment with different $\sigma_{\rm \chi N}^{\rm SI}$. }
	\label{fig3_exclusion} 
\end{figure}

With the XENONnT public dataset, no significant signal is observed. The one-sided 90\% confidence level (CL) constraints on ($\sigma_ev_\chi$) with both V and A operators for $m_\chi$ $\in$ (1$\times$10$^{-5}$--1) keV are obtained, adequately encompassed by the XENONnT dataset. The results are shown in Fig.~\ref{fig3_exclusion}. 
The numerical value of the constraints on ($\sigma_ev_\chi$) is inversely proportional to $\sigma^{\rm SI}_{\rm \chi N}$ at corresponding $m_\chi$. Owing to the absence of experimental constraints on $\sigma^{\rm SI}_{\rm \chi N}$ for $m_\chi$ $\in$ (1$\times$10$^{-5}$--1) keV, the $\sigma^{\rm SI}_{\rm \chi N}$ is set to 1 $\times$ 10$^{-31}$ cm$^2$ from a previous study on CRDM~\cite{novel} for a conservative evaluation. Nevertheless, this does not affect the span of $m_\chi$ for which the constraints can be derived. Our results exclude the fermionic DM-electron absorption cross section ($\sigma_e v_\chi$) mediated by V (A) operator from 8.0$\times$10$^{-42}$ (2.7$\times$10$^{-42}$) -- 2.8$\times$10$^{-34}$ (9.3$\times$10$^{-35}$) cm$^2$ for $m_\chi$ $\in$ (1$\times$10$^{-5}$--1) keV. This particular region remains unexplored in previous studies and our results reach the lowest $m_\chi$ to date among the DMDD experiments utilizing the ionizing signals from DM interactions within the detectors.

\emph{Conclusion.}—
We demonstrate that the subdominant while energetic component of DM boosted by CR can impose a novel constraint on ($\sigma_e v_\chi$) for a range of $m_\chi$ which was previously deemed inaccessible.
\par
In this work, we explored the parameter space on ($m_\chi$, $\sigma_e v_\chi$) of ultra-light fermionic DM-electron absorption induced by V and A operators with the consideration of boost from CR for the first time. Employing the 1.16 tonne$\cdot$year XENONnT public dataset, the one-sided 90\% CL constraints on ($\sigma_ev_\chi$) for $m_\chi$ $\in$ (1$\times$10$^{-5}$--1) keV are derived. Constraints have been derived on DM at the smallest $m_\chi$ currently accessible among DMDD experiments using ionizing signals from DM interactions within detectors.

\bibliography{main}

\end{document}